# X-ray luminescence computed tomography using a focused X-ray beam

Wei Zhang, Michael C. Lun, Alex Anh-Tu Nguyen, Changqing Li[*]

School of Engineering, University of California, Merced, Merced, CA 95343, USA

*cli32@ucmerced.edu

# Abstract

Due to the low X-ray photon utilization efficiency and low measurement sensitivity of the electron multiplying charge coupled device (EMCCD) camera setup, the collimator based narrow beam X-ray luminescence computed tomography (XLCT) usually requires a long measurement time. In this paper, we, for the first time, report a focused X-ray beam based XLCT imaging system with measurements by a single optical fiber bundle and a photomultiplier tube (PMT). An X-ray tube with a polycapillary lens was used to generate a focused X-ray beam whose X-ray photon density is 1200 times larger than a collimated X-ray beam. An optical fiber bundle was employed to collect and deliver the emitted photons on the phantom surface to the PMT. The total measurement time was reduced to 12.5 minutes. For numerical simulations of both single and six fiber bundle cases, we were able to reconstruct six targets successfully. For the phantom experiment, two targets with an edge-to-edge distance of 0.4 mm and a center-to-center distance of 0.8 mm were successfully reconstructed by the measurement setup with a single fiber bundle and a PMT.

## 1. Introduction

X-ray luminescence computed tomography (XLCT) is an emerging hybrid imaging modality, in which high energy X-ray photons are used to excite phosphors that emit optical photons to be measured for optical tomographic imaging [1]. XLCT has the potential to be a powerful tool in molecular imaging because it is possible for XLCT to combine the high spatial resolution of X-ray imaging and the high sensitivity of optical imaging. So far, several XLCT systems have been designed and studied. Pratx *et al*. proposed the idea of hybrid X-ray luminescence optical tomography and built the first prototype system [1]. They proved that XLCT imaging was able to reconstruct the distribution of phosphor particles by using a selective X-ray beam scanning scheme [2]. Li *et al*. experimentally demonstrated that XLCT was capable of obtaining high spatial resolution by using collimated X-ray beams for deep targets [3]. Zhang *et al.* proposed a multiple pinhole based XLCT design in which multiple X-ray beams were used to scan objects simultaneously to reduce the data acquisition time. An edge-to-edge distance of 0.6 mm has been achieved by using this multiple pinhole XLCT system [4]. Chen *et al.* proposed a cone beam XLCT design which improves the data acquisition time at the cost of a degraded spatial resolution [5]. Additionally, Liu *et al.* applied a cone beam based XLCT for small animal imaging [6]. Recently, Lun *et al.* have reported that a phosphor target with the concentration as low as 0.01 mg/mL at a scanning depth of 21 mm could be reconstructed successfully and the reconstructed target size varied by less than 4% for different scanning depths between 6 mm and 21 mm [7]. They have also shown that the measurement sensitivity of XLCT is at least 100 times better than a typical CT scanner in imaging the X-ray excitable phosphor target in deep turbid media. All these reports have demonstrated that XLCT is a promising tool for *in vivo* small animal imaging.

One of the main reasons why XLCT has not yet been adopted by the molecular imaging community is its long scanning time [8]. The conical X-ray beam based XLCT could reduce the scan time, however, the spatial resolution was degraded to several millimeters since the X-ray beam's position can no longer be used as structural guidance in the reconstruction [5]. There are several approaches to improve the scanning speed of the pencil beam based XLCT which include approaches such as using a higher flux X-ray beam, brighter nanophosphors, and more sensitive optical photon detectors [8].

One way to improve the X-ray photon flux in a collimated X-ray beam is to use a powerful X-ray tube, but it is not efficient because most X-ray photons are absorbed by the collimator. X-ray optics has been studied to focus X-ray beams into a small spot with a large X-ray flux, which has been used successfully in X-ray fluorescence tomography imaging [9]. Cong *et al.* proposed a focused beam based XLCT, in which 50 times more X-ray photon intensity was obtained by delivering focused X-ray beams onto a spot with a diameter of 50 µm, and they have demonstrated the feasibility of their concept with numerical simulations [10].

We reported the first experimental demonstration of a polycapillary lens based XLCT imaging system and verified its feasibility with numerical and phantom experiments [11]. To investigate the polycapillary lens based XLCT approach further, here we performed a systematic study of the focused X-ray beam based XLCT. In this paper, we present an experimental XLCT imaging system that used a polycapillary lens to focus the X-ray beam, a single optical fiber bundle to collect the emitted optical photons, and a highly sensitive PMT to measure the optical signal. The feasibility of the design was demonstrated through both numerical simulations and phantom experiments. We compared the X-ray photon flux between a focused X-ray beam and a collimated X-ray beam. To investigate the effect of the X-ray lens, we measured the X-ray energy spectrum for the X-ray tube with and without the X-ray lens. For X-ray dose study, we measured the radiation dose of this XLCT system.

Another way to reduce the XLCT scanning time is to utilize optical photon detectors with higher sensitivity in the XLCT system. Currently, all the XLCT systems have been equipped with an EMCCD camera to measure emitted photons. With a much higher sensitivity than the EMCCD cameras, photomultiplier tubes (PMTs) are preferred, especially for the detection of weak optical signals as in XLCT. As a high gain device, PMTs have already been used widely as detectors in optical imaging. Here a PMT was used as the optical detector in XLCT to achieve faster data acquisition.

This paper is organized as follows. In Section 2, we introduce the focused X-ray beam based XLCT system configuration, the reconstruction method, the numerical simulation setup and the experimental measurement setup. Section 3 describes the numerical simulation and phantom experimental results. Finally, in Section 4, we discuss our results and conclude the paper.

**2. Methods**

*2.1 Focused X-ray beam based XLCT imaging system*

Based on previous studies, we proposed a focused X-ray beam based XLCT imaging system. The schematic of the new XLCT system is plotted in Fig. 1. An X-ray tube (Polycapillary X-Beam Powerflux, XOS, NY; Target metal: Molybdenum (Mo)) was utilized to generate X-ray photons up to the maximum energy of 50 kVp at a tube current of 1 mA. The output X-ray beam was focused by a 79.2 mm long, 45 mm output focal distance polycapillary X-ray lens with a focal spot size of 100 µm. Phantoms were placed 45 mm away from the output of the X-ray lens and placed on a motorized rotation stage (B4872TS-ZR, Velmex, Inc., NY) which was mounted on a motorized linear stage (MB2509Q1J-S3, Velmex, Inc., NY). The passed X-ray beam was detected by an X-ray detector (Shad-o-Box 1024, GOS scintillator screen, Rad-icon Imaging Corp., CA), which had a detection area of 49.2×49.2 mm$^2$ consisting of a 1024×1024 pixel photodiode array sensor with a 48 µm pixel size. The X-ray detector measured the intensity of the focused X-ray beam, from which the phantom boundary was detected. The emitted optical photons from the phantom side surface were collected by a 2 meters long fiber bundle with an aperture diameter of 3 mm. The fiber bundle was 2 mm away from the phantom surface so that we could adjust the scanning depth with the jack [7]. The fiber bundle was fixed by a mount frame that moved and rotated with the phantom. A fan-cooled PMT (H7422P-50, Hamamatsu, Japan) driven by a high voltage source (C8137-02, Hamamatsu, Japan) measured the optical photons from the fiber bundle. The electronic signal from the PMT was further amplified by a broadband amplifier (SR445A, Stanford Research Systems, CA) with a gain of 25. Then, a low pass filter (BLP-10.7+, cutoff frequency 11 MHz, Mini-circuits) was used to reduce the high frequency noise. The amplified and filtered signal was finally acquired and displayed by a high-speed oscilloscope (MDO-3014, Tektronix, OR). The whole system except the PMT, the amplifier and the oscilloscope was fixed on an optical bench and placed in an X-ray shielding and light tight cabinet. All the devices were controlled by a lab-made program written with C++ in the Visual Studio® development environment.

*2.2 Comparison of X-ray photon flux between a focused X-ray beam and a collimated X-ray beam*

To evaluate how much improvement in X-ray flux we can obtain by the polycapillary lens, we designed a comparison test between a focused X-ray beam and a collimated X-ray beam. We evaluated the X-ray photon flux by measuring the intensity of the emitted luminescence photons acquired by the EMCCD camera (C9100-3, Hamamatsu, Japan) from the top surface of a cylindrical phantom embedded with a phosphor target while an X-ray beam from a focused or collimated X-ray tube excited the phosphor target. The setups for the test are shown in Fig. 2. In the first setup as plotted in Fig. 2(a), we used the XOS X-Beam X-ray tube (Mo target), in which a polycapillary lens was applied and an X-ray beam with a focal spot size of 0.098 mm was generated to excite the target. The X-ray tube current was set to be 1 mA and the tube voltage was 50 kVp. In the second setup as plotted in Fig. 2(b), like our previous work [4], an X-ray tube (93212, Oxford Instruments; Tungsten (W) target) was used. A collimator was employed to generate a 1.0 mm diameter pencil beam that excited the target. The X-ray tube current was set to be 2 mA and the tube voltage was 50 kVp. In both setups, the same cylindrical phantom embedded with a 4.6 mm diameter cylindrical phosphor target ($Eu^{3+}$-doped gadolinium oxysulfide, $GOS:Eu^{3+}$) at a concentration of 1 mg/mL was used. The EMCCD camera exposure time for the focused beam and the collimated beam was 0.1 and 2 seconds, respectively.

We had to use different X-ray tubes and measurement parameters in this comparison study because the polycapillary lens is fixed and could not be removed from its X-ray tube. To compensate the effects of different settings, we normalized the measurements to the X-ray tube power, the X-ray beam diameter, and the exposure time. After acquiring the EMCCD camera images of phantom top surface for both the focused X-ray beam case and the collimated X-ray beam case, background noise was firstly subtracted from the acquired EMCCD camera images. Then the images were normalized to the X-ray tube power, the X-ray beam size, and the exposure time as:

$$\begin{cases} Nf_i = f_i/(0.1s)/(1mA * 50kVp)/(\pi * 0.1 * 0.1 mm^2) \\ Nc_i = c_i/(2s)/(2mA * 50kVp)/(\pi * 1 * 1 mm^2) \end{cases} \quad i = 1, 2, \dots, M \quad (1)$$

where $f_i$ and $c_i$ are measurements at the $i^{th}$ pixel in the EMCDD images for the focused X-ray beam case and the collimated X-ray beam case, respectively. $Nf_i$ and $Nc_i$ are the normalized pixel values of the corresponding images. $M$ is the number of the image pixels.

The ratio between the total intensity of the focused X-ray beam and the total intensity of the collimated X-ray beam was calculated as:

$$Ratio = \sum_{i=1}^{M} Nf_i / \sum_{i=1}^{M} Nc_i \qquad (2)$$

*2.3 Energy spectra and beam size of the focused X-ray beam*

To investigate how the X-ray lens affects the X-ray energy spectrum, we have measured the X-ray energy spectrum for the X-ray tube with the lens by using a thermoelectrically cooled cadmium telluride (CdTe) detector (X-123 CdTe, Amptek Inc., Bedford, MA) for the tube voltages of 30, 40 and 50 kVp, respectively. The X-ray detector module includes a preamplifier with pile-up rejection, a digital pulse processor and a multichannel analyzer (MCA) (PX4, Amptek Inc., Bedford, MA). The X-ray tube vendor (XOS, NY) measured the X-ray energy spectrum for the X-ray tube without the X-ray lens by using a silicon drift detector (XR-100SDD, Amptek Inc., Bedford, MA) for the tube voltages of 30, 40 and 50 kVp, respectively. For the energy spectrum measurement without the X-ray lens, the X-ray tube current was 0.3 mA and the exposure time was 100 seconds. Additionally, the detector used a pinhole of 0.5 mm and was 487 mm away from the X-ray tube. For the energy spectrum measurement with the X-ray lens, we also took measurements at the tube current of 0.3 mA and used 100 seconds of exposure time. The X-ray spectrometer (X-123 CdTe) used a 0.1 mm pinhole and was positioned 200 mm away from the lens.

Gafchromic EBT3 films were mounted on a linear stage to measure the size and intensity of the focused X-ray beam at different distances. The step size of the linear stage was 3 mm with 8 steps. We measured the X-ray beam size and intensity at a tube current of 0.25 mA and with varying tube voltages (20, 30, 40, and 50 kVp). The exposure time of the film for each linear step was 10 seconds. After being exposed, all the films were scanned by a high resolution scanner (Epson Expression 11000XL). The intensity and the diameter of the focused X-ray beams were calculated from the pictures captured by the scanner by analyzing the exposed spot size and intensity.

*2.4 Measurement of radiation dose in XLCT*

We performed a dose measurement experiment as shown in Fig. 3. The X-ray dose was measured using an Accu-Dose system (Radcal, Monrovia, CA) with a general purpose in-beam ion chamber (10X6-6, Radcal). The active component of the ion chamber head has a diameter of 25 mm. The phantom was 44 mm in diameter and contained a central hole to fit the ion chamber head and was composed of 1% Intralipid and 2% Agar. The phantom was placed on the rotary stage mounted on the linear stage. The ion chamber was fit into the phantom center. We then performed a scan that was the same as the scan in the following phantom experiment. We used 125 linear scan steps with a step size of 0.2 mm, 6 angular projections with an angular step size of 30 degrees and a measurement time of 1 second per linear scan step.

*2.5 Numerical simulation setup*

To validate our proposed focused X-ray beam based XLCT design with measurements by optical fiber bundles, we have performed two numerical simulation cases both using a 6 target phantom while taking measurements using one and six optical fiber bundles, respectively. For both simulation studies, we used a 10 mm long cylindrical phantom with a diameter of 12.8 mm. The optical properties of the phantom were set to be $\mu_a = 0.0072\ mm^{-1}$ and $\mu_s' = 0.72\ mm^{-1}$ [4]. The X-ray attenuation coefficient was $\mu_x = 0.0214\ mm^{-1}$. All the six targets had a diameter of 0.2 mm and a height of 6 mm and were embedded in the phantom. The positions of the targets are shown in Fig. 4, from which we can see that the target center-to-center distance was 0.4 mm. For both numerical studies, we set the phosphor particle concentration to be 1 mg/mL in targets and 0 mg/mL (no phosphors) in the background.

The fiber bundles were placed at 3 mm under the phantom top surface. The relative positions of fiber bundles to the phantom were fixed. During experiments, the fiber bundles and the phantom translated and rotated together. In the six fiber bundle detection case, the fiber bundles were distributed uniformly with an angular step of 30 degrees as shown in Fig. 4. In the single fiber bundle detection case, only the #4 fiber bundle was used to collect emitted photons. For both simulation cases, we used a focused X-ray beam to scan the phantom at a depth of 5 mm. The focused X-ray beam diameter and the linear scan step size were set to be 100 μm. For both numerical simulations, we used six angular projections with an angular step size of 30 degrees. The numerical measurements were generated from the forward model in which the phantom was discretized by a finite element mesh with 26,638 nodes, 153,053 tetrahedral elements and 11,456 face elements. Finally, 10% Gaussian noise was added to the numerical measurements.

In the focused X-ray beam XLCT, the shape of the focused X-ray beam was a dual-cone. In this paper, we took the true beam shape into consideration. As described in the above section, we measured the focused X-ray beam size and intensity and found that the focused X-ray beam was dual-conical. The beam diameter changed linearly and the beam intensity attenuated exponentially as the distance from the collimator increased [4]. We set the focal distance of the X-ray lens to be 4.5 mm. The beam diameter at position $\vec{r}$ can be expressed as:

$$d(\vec{r}) = \begin{cases} 0.2 - L(\vec{r})/45, & L(\vec{r}) \leq 4.5 \\ L(\vec{r})/45, & L(\vec{r}) > 4.5 \end{cases} \quad (3)$$

where $L \in [0, 12.8]$ is the distance from one side to another side of the phantom.

In numerical simulations, we adopted a normalized X-ray beam intensity. Therefore, the X-ray intensity at the entry to the phantom ($T_0$) was assumed to be equal to 1. The X-ray attenuation coefficient was $\mu_x = 0.0214\ mm^{-1}$ in the phantom. Then the X-ray intensity along the X-ray beam in the phantom is given by the following equation:

$$T(\vec{r}) = exp(-0.0214 \times L(\vec{r})) \tag{4}$$

For both numerical simulation cases, we have included the true dual-conical X-ray beam geometry in the forward model.

*2.6 Phantom experimental setup*

We performed a phantom experiment using a solid cylindrical phantom embedded with two targets as shown in Fig. 5. The background phantom was 40 mm long and 25 mm in diameter and was composed of 1% $TiO_2$ and 200 ml resin. For the targets, we used two glass capillary tubes (Capillary Tubes 1000-800/12, Drummond Scientific Company) placed side by side, as shown in Fig. 5(b), with an inner diameter of 0.4 mm and a wall thickness of 0.2 mm. The tubes were filled with a solution of 1% Intralipid, 2% Agar, and 10 mg/ml GOS:$Eu^{3+}$ (UKL63/UF-R1, Phosphor Technology Ltd). As shown in Fig. 5(c), the center positions of targets were at (-0.4 mm, -6.5 mm) and (0.4 mm, -6.5 mm). During the phantom experiment, the X-ray detector was used to determine the phantom boundary for beginning measurement by examining the X-ray beam intensity changes. The X-ray beam scanning depth was 5 mm during the experiment and a single optical fiber bundle was placed 10 mm below the top surface of the phantom to collect the emitted optical photons from the side surface as shown in Figs. 5(c) and 5(d). We acquired measurements from 6 angular projections, using an angular step size of 30º. To scan the entire diameter of the phantom, 125 linear steps with a step size of 0.2 mm were required at each angular projection. During data acquisition, the fan-cooled PMT was operated with a control voltage of 0.75 V and the amplifier was operated at a gain of 25. For each linear scan step, the oscilloscope measurement time was set to be 1 second. The current measurement time is 6×125×1 seconds or 12.5 minutes. Finally, during the entire experiment, the X-ray tube was operated at 30 kVp and 0.5 mA.

*2.7 Evaluation criteria*

Three criteria were used to evaluate the quality of the reconstructed XLCT images, as described in Ref. 4:

Target Size Error (TSE): This criterion is defined as the target diameter error ratio between the reconstructed target and the true target and is given by the following equation:

$$TSE = \frac{|D_r - D_t|}{D_t} \times 100\% \qquad (5)$$

where $D_r$ and $D_t$ are the diameters of reconstructed and true target, respectively. $D_r$ is calculated from the cross target profile plot by using the full width at half maximum (FWHM) approach, in which we measured the width at the half of the maximum.

Center-to-center Distance Error (CDE): For multiple target imaging, we define CDE as the distance error ratio between the reconstructed targets and the true targets and is given by:

$$CDE = \frac{|Dist_r - Dist_t|}{Dist_t} \times 100\% \qquad (6)$$

where $Dist_r$ and $Dist_t$ are the center-to-center distances (CtCD) between the reconstructed targets and the true targets, respectively. $Dist_r$ is also calculated from the cross target profile plot by using the FWHM approach.

Dice Similarity Coefficient (DICE): DICE is used for comparing the similarity between the reconstructed and true targets and is given by:

$$DICE = \frac{2 \times |ROI_r \cap ROI_t|}{|ROI_r| + |ROI_t|} \times 100\% \qquad (7)$$

where $ROI_r$ is the reconstructed region of interest that is defined to be the pixels whose intensities are higher than 10% of the maximum of the normalized reconstructed intensity, and $ROI_t$ is the true target region. Generally, the closer DICE is to 100%, the better.

## 3. Results

*3.1 X-ray photon flux study*

Figs. 6(a) and 6(b) show the normalized images for the focused X-ray beam and the collimated X-ray beam cases, respectively. We plotted the profile plots along the green lines in Fig. 6(a, b) as shown in Fig. 6(c). From Figs. 6(a) and 6(b), we found that the ratio of the maximum photon intensities between the focused beam case and the collimated beam case was as large as 2013. We also calculated the intensity summation of the entire phantom top surface for each image and we found that the total intensity of the focused beam case was 1200 times larger than that of the collimated beam case. Therefore, we can conclude that the focused X-ray beam can deliver 1200 times more X-ray photon density (X-ray photon number per beam volume within the fluorophore) than the collimated X-ray beam.

*3.2 Energy spectra, beam size and intensity of the focused X-ray beam*

The measured X-ray energy spectra for different tube voltages (30, 40, and 50 kVp) are plotted in Fig. 7(a) for the X-ray tube without the lens and Fig. 7(b) for the X-ray tube with the lens. The vertical axis indicates the photon counting number recorded by the X-ray photon detectors. From both plots, we see there are two energy peaks at 17.5 keV and 19.5 keV corresponding to the characteristic X-ray photons of Mo. We also observed higher photon number ratio in the low energy range when we used the X-ray lens, which is reasonable because low energy X-ray photons are easier to be focused. To quantify the analysis, we have calculated the X-ray photon count ratio of the X-ray photons with energies less than the peak of 17.5 keV to all X-ray photon number for the 50 kVp case. The ratios were 58.7% and 70.2% for the X-ray tube without and with the lens, respectively. We found that the lens increased low energy X-ray photon ratio about 11.5%.

We measured the diameter and the intensity of the focused X-ray beam at different settings (X-ray tube voltage: 20, 30, 40, and 50 kVp; X-ray tube current: 0.25 mA). For simplicity, we only plotted the result measured when the X-ray tube voltage was 30 kVp, which was the same voltage used in the experimental study. Fig. 8 shows the raw film images (Fig. 8(a)), the measured X-ray beam diameter (Fig. 8(b)), the profile plots across the X-ray beam (Fig. 8(c)), the maximum X-ray intensity (Fig. 8(d)), and the averaged intensity (Fig. 8(e)) for the focused X-ray beam. The measured X-ray beam diameter demonstrates that the focused X-ray beams are dual-conical and the beam diameter changes as the distance increases. As seen in Fig. 8(b), the smallest X-ray beam diameter is 98 micrometers, slightly less than 100 micrometers, at the focal spot of 45 mm from the polycapillary lens. The intensity curves show that beyond the focal spot, the X-ray beam intensity attenuates exponentially. We observed that as the X-ray tube voltage increases, the X-ray beam diameter also increases. From Fig. 8, we see that the scanned object should be 38 mm away from the lens to have small X-ray beam diameter.

*3.3 Radiation dose in XLCT*

The total accumulated ionized X-ray radiation was 7.38 R. Using an f-factor (conversion of exposure in air to absorbed dose in muscle at a diagnostic X-ray energy of 10 keV) of 0.93 rad/R or (cGy/R), we calculate the absorbed dose to be 68.634 mSv or 6.8634 cGy.

*3.4 Results of numerical simulations*

The scanned transverse section was discretized with a 2D grid having a pixel size of $25 \times 25\ um^2$. The system matrix generated with the finite element mesh was interpolated to the fine 2D grid. During the reconstruction, the $L^1$ regularization method was applied using a majorization-minimization (*MM*) algorithm reconstruction framework to solve the optimization problem [13-15]. Fig. 9 shows the reconstructed XLCT images for numerical simulations of six targets with one fiber bundle case and six fiber bundle case, respectively.

From Fig. 9, we see that all six targets have been reconstructed at the correct positions with an acceptable shape by using $L^1$ regularized *MM* algorithm with measurements at six angular projections. For simplicity, we only drew the normalized profile plot across the middle row targets in Fig. 9 and calculated the image quality metrics as shown in Table 1. We can see that the six fiber bundle method has a better performance in terms of TSE and CDE than the one fiber bundle method as shown in Table 1. The image quality metrics degrades slightly when using one fiber bundle. Compared with six fiber bundle case, measurement data obtained from one fiber bundle is sufficient to reconstruct a good XLCT image, although the six fiber bundle case improves the reconstruction result slightly.

*3.5 Results of phantom experiment*

The $L^1$ regularized *MM* algorithm was again used to reconstruct XLCT images for the phantom experiments as in the numerical simulations. For the XLCT reconstruction, the phantom was discretized by a finite element mesh with 26,638 nodes, 153,053 tetrahedral elements and 11,456 face elements. The reconstructed transverse section was also discretized with a 2D grid with a pixel size of $50 \times 50\ um^2$ for reconstruction. The measured X-ray beam size models were applied in the XLCT reconstruction.

Fig. 10 shows the results of the phantom experiment. In Fig. 10(b), the reconstructed XLCT image is plotted. Fig. 10(c) shows the zoomed in target region (green box in Fig. 10(b)) where the green circles represent the true targets' size and locations which are determined from the microCT image of our phantom given in Fig. 10(a). From the reconstructed image, we see that the two targets were reconstructed successfully and have been clearly resolved. To further analyze the reconstructed XLCT image quantitatively, we have calculated the image quality metrics as shown in Table 2. From the dotted blue line in Fig. 10(c), a normalized line profile is plotted in Fig. 10(d). From the full width half maximum, we calculated a reconstructed target size of 0.45 mm with an error of 12.5%. In addition, the center-to-center distance is 0.75 mm with an error of 6.25% and the DICE was evaluated to be 80.0%. Based on our results, multiple targets can be successfully reconstructed using the proposed XLCT system.

## 4. DISCUSSION AND CONCLUSIONS

In this study, we have, for the first time, used a polycapillary lens to focus the output X-ray beam for a higher X-ray photon flux during the XLCT imaging. We have generated an X-ray beam with a diameter as small as 98 micrometers at the focal spot of the lens. We compared the photon flux between the focused X-ray beam and a collimated 1 mm diameter X-ray beam and found 1200 times X-ray photon flux density increase by the focused X-ray lens. In previous work, it took 130 minutes to scan a 12.8 mm diameter phantom by using a superfine X-ray beam of 0.175 mm in diameter [16]. In this study, the total measurement time was 12.5 minutes to scan a 25 mm diameter phantom by using a focused X-ray beam. Due to the improvement on X-ray flux density we can dramatically reduce the required scan time for XLCT imaging. From Fig. 6, we can see that with a 0.1 s exposure time for the 0.1 mm focused X-ray beam (6(a)), we have a much brighter EMCCD camera image than by using a 2 second exposure time with a 1 mm collimated X-ray beam (6(b)). In the future, we would like to design an XLCT imaging system to perform the linear scan in a continuous scan mode if a PMT detector is used, in which each angular scan will take less than 10 seconds. And we will complete the total scan in less than 60 seconds if 6 angular projections are needed.

The fiber bundle position is a factor in the XLCT imaging setting. To investigate how the fiber bundle position affects the XLCT reconstruction, we have performed several numerical simulations. All the settings are the same with one fiber bundle as one detector except the different fiber bundle position. We placed the fiber bundle at different angles (30, 45, 60, 90, 270, 300, 330 and 360 degrees) as shown in Fig. 11. The reconstructed images are shown in Fig. 12, from which we see that the reconstructed image results become better as the fiber bundle moved close to the six targets. The case with the fiber bundle at 270 degrees has the best result. When the fiber bundle was at 90 degrees (the furthest position from the targets), the reconstruction result is still acceptable with all six targets reconstructed successfully. In section 2.5, we placed the fiber bundle at 90 degrees in numerical simulations and 360 degrees in the phantom experiments in section 2.6. The bottom left target has barely been reconstructed when the fiber bundle was placed at 30, 45, or 60 degrees. This is reasonable because multiple targets were excited simultaneously and we only had one detector with 6 angular projections. In the future, we can have measurements at more angular projections or more fiber bundles as detectors to overcome this issue.

Although a PMT can only measure the optical intensity at one spot, our numerical simulation and phantom experiment have demonstrated that sparse sampling with only one fiber bundle is sufficient to reconstruct complex targets deeply embedded in turbid media. Our numerical simulation results also show that more measurements can improve the XLCT image quality. In the future, we will use more fiber bundles with more PMTs to achieve better XLCT images.

One concern is that the energy of most X-ray photons in the focused X-ray beam is within the range from 15 keV to 20 keV as shown in Fig. 7, which might mean that the focused X-ray beam cannot penetrate large objects. We have performed an attenuation measurement and found that there are sufficient X-ray photons passed a 2 cm thick agar phantom, which implies that our focused X-ray beam is appropriate for imaging mice-size objects.

It is ideal if we can find a way to count the X-ray photon numbers directly. The Gafchromic EBT3 films might be used to count the X-ray photon number. However, it works well only if the X-ray photons have the same energy. The ultimate goal in this project is to have an X-ray beam to excite the phosphor target for the brightest luminescence signals. Thus, we used the EMCCD camera to measure and compare the luminescence intensity on the top surface when the same target was excited by the collimated X-ray beam and the focused X-ray beam. The luminescence intensity is proportional to the X-ray photon number if the X-ray photons have the similar energies. We found the luminescence intensity from the focused 0.1 mm X-ray beam excitation was 12 times larger than that from the 1.0 mm collimated X-ray beam. Considering the 10 times difference of the beam diameter (or 100 times difference of beam section area), the X-ray flux density in the focused X-ray beam is 1200 times larger than the focused X-ray beam. It is worth noting that the collimated X-ray beam was from a Tungsten (W) X-ray tube which has higher X-ray energy peaks than the Molybdenum (Mo) X-ray tube used for the focused X-ray beam. Therefore, the X-ray photon flux density in the focused X-ray beam was more than 1200-fold of that in the collimated X-ray beam.

The measured X-ray dose with the focused X-ray beam per XLCT scan is 68.634 mSv, which is 46 times higher than what we reported in [4]. It is reasonable because the focused X-ray beam has much greater X-ray photon flux. We will use the X-ray shutter inside the X-ray tube to reduce the X-ray radiation dose in XLCT imaging in the future studies. In the future, we will scan the object in a continuous scan mode with much shorter scan time which will reduce the dose substantially.

Another challenge in this study is to find initial and final X-ray beam positions that correspond to the actual starting and ending points for each projection. The X-ray detector can help by detecting the changes in the measured X-ray beam attenuation, however, the intensity changes are not always obvious or not sharp for the first and last positions of the scan which leads to errors in mapping the X-ray beam positions, which caused some errors in our XLCT reconstructed images (Fig. 10(c)). In the future, we will design a rotating X-ray beam and fix the scanned object so that the field of view will be fixed.

For the phantom experiments, we have set the linear scan step size to be 200 µm that is equal to the average dual cone diameter. For the numerical simulations, we have used a linear scan step size to be 100 µm, close to the minimum diameter of the dual cone X-ray beam. For both cases, we have reconstructed the targets very well, which indicates that the linear scan step size can be larger than the minimum X-ray beam diameter. In future study, we will use a continuous scan mode with the PMT as detector. We can select any linear scan step size in the post-processing of the measurements.

In sum, we have built a focused X-ray beam based XLCT imaging system for the first time. We can perform sparse sampling with a single optical fiber bundle and a PMT. The X-ray flux in the focused X-ray beam is 1200 times larger than that of a collimated X-ray beam. Our numerical simulation and phantom experimental results have demonstrated the feasibility of the proposed focused X-ray beam based XLCT imaging system.

## ACKNOWLEDGEMENTS

This work is supported in part by Grants (R03 EB022305) from the National Institute of Health (NIH) and Start Up fund from UC Merced. The authors thank Prof. Simon R. Cherry from University of California, Davis for lending us the microCT setup and XOS for assisting in the X-ray tube spectrum measurements.

## CONFLICT OF INTEREST DISCLOSE

The authors have no COI to report.

**FIGURE CAPTIONS:**

**Fig. 1** Schematic of the focused X-ray beam based XLCT imaging system.

**Fig. 2** The setups of X-ray photon flux comparison for the focused X-ray beam (a) and the collimated X-ray beam (b).

**Fig. 3** The schematic design (left) and a photo (right) of the X-ray dose measurement setup.

**Fig. 4** The phantom geometry and fiber bundle positions for numerical simulations with six targets.

**Fig. 5** (a) White light picture showing the side view of the solid phantom (left), the targets (middle) and a penny (right) as reference. (b) Top view of the solid phantom, where two capillary tubes as targets were placed inside the hole of the solid phantom. (c) The phantom geometry used for the experiment, where two capillary tubes are the targets. (d) The focused X-ray beam based XLCT system setup.

**Fig. 6** Normalized phantom top surface images acquired by the EMCCD camera (a) with the 0.1 mm focused X-ray beam and (b) with the 1 mm collimated X-ray beam. (c) Profile plots along the green lines in (a) and (b).

**Fig. 7** Measured X-ray photon energy spectra of the X-ray tube without the lens (a) and with the lens (b) for the X-ray tube voltages of 30, 40 and 50 kVp, respectively.

**Fig. 8** Measurement of focused X-ray beam diameter and intensity: (a) Original film images obtained at different distances. (b) X-ray beam diameter at different distances from the polycapillary lens. (c) Profile plot across the X-ray beam at different distances. (d) Maximum X-ray intensity at different distances. (e) Mean X-ray intensity at different distances.

**Fig. 9** Reconstructed XLCT images, zoomed in regions and profile plots for numerical simulations with six targets. (a) Reconstructed results with data from one fiber bundle; (b) Reconstructed results with data from six fiber bundles.

**Fig. 10** (a) A transverse section from the reconstructed microCT image of the phantom with two targets. (b) Reconstructed XLCT image. (c) The zoomed in image of the reconstructed image. (d) The profile plot across the two targets. The green square in (b) indicates the zoomed in region. The green line in (c) indicates the exact target size and position. The blue dotted line in (c) indicates the profile location.

**Fig. 11** Phantom geometry and fiber bundle positions for the numerical simulation studies on the effect of fiber bundle position in XLCT imaging.

**Fig. 12** Reconstructed XLCT images for the numerical simulations with different fiber bundle positions. The angle indicates the single fiber bundle position.

**Table Captions:**

**Table 1.** Quantitative imaging quality metrics for the numerical simulations with one and six fiber bundles, respectively.

**Table 2.** Quantitative imaging quality metrics for the phantom experiment with two targets.

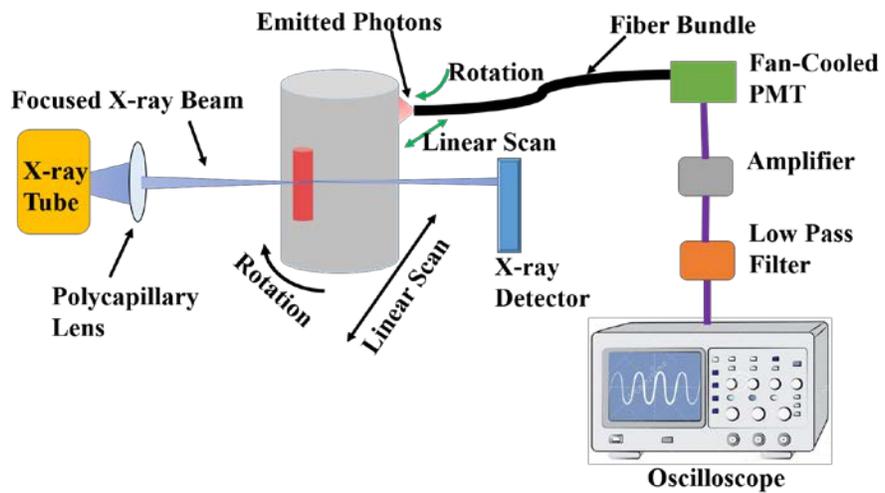

Fig. 1

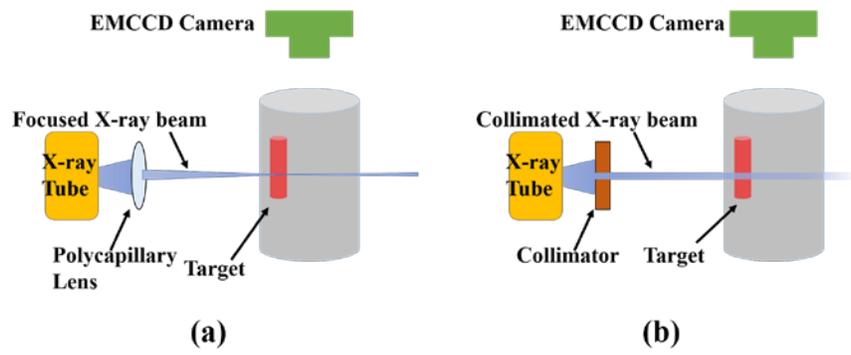

Fig. 2

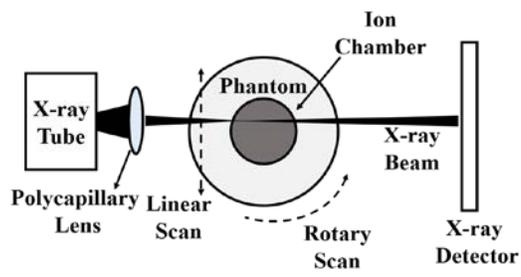 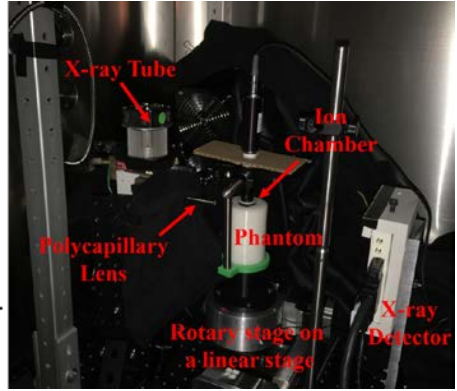

Figure 3

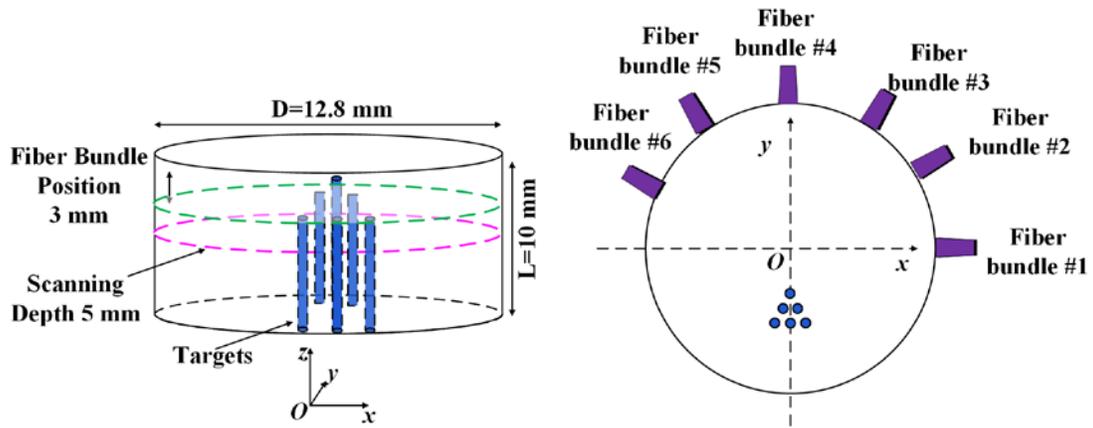

Figure 4

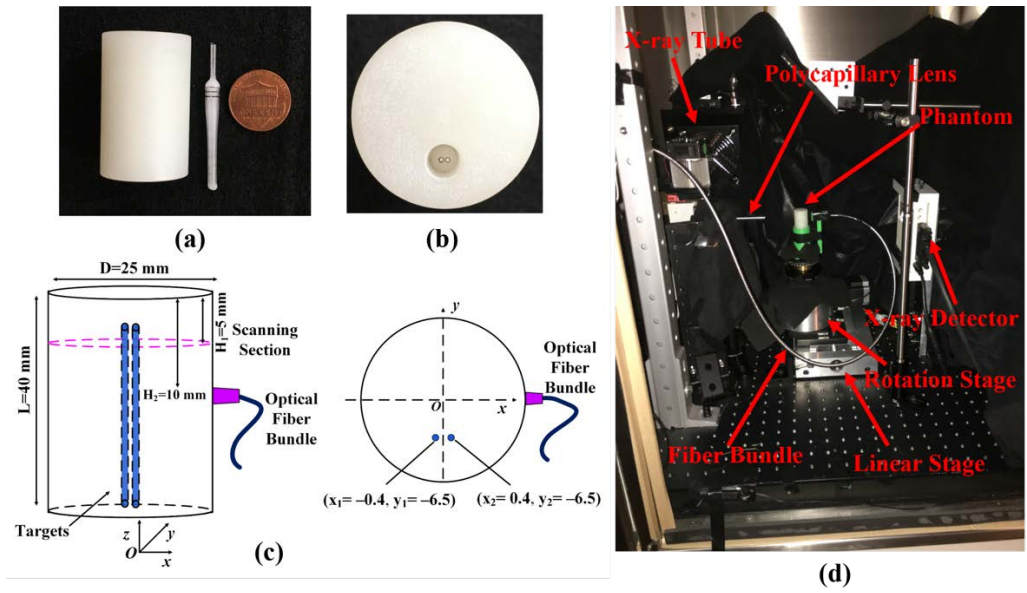

Figure 5

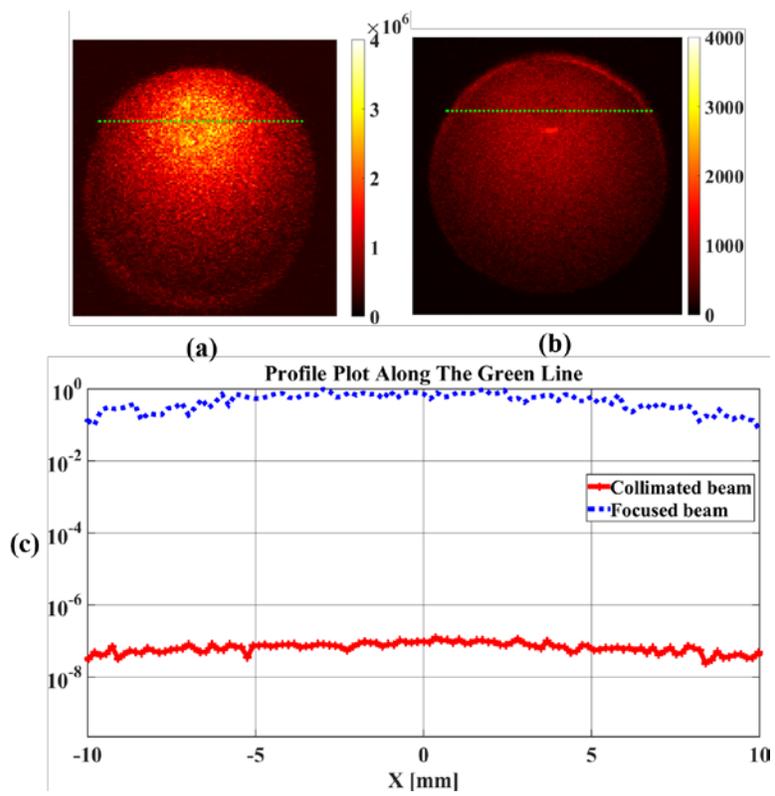

Figure 6

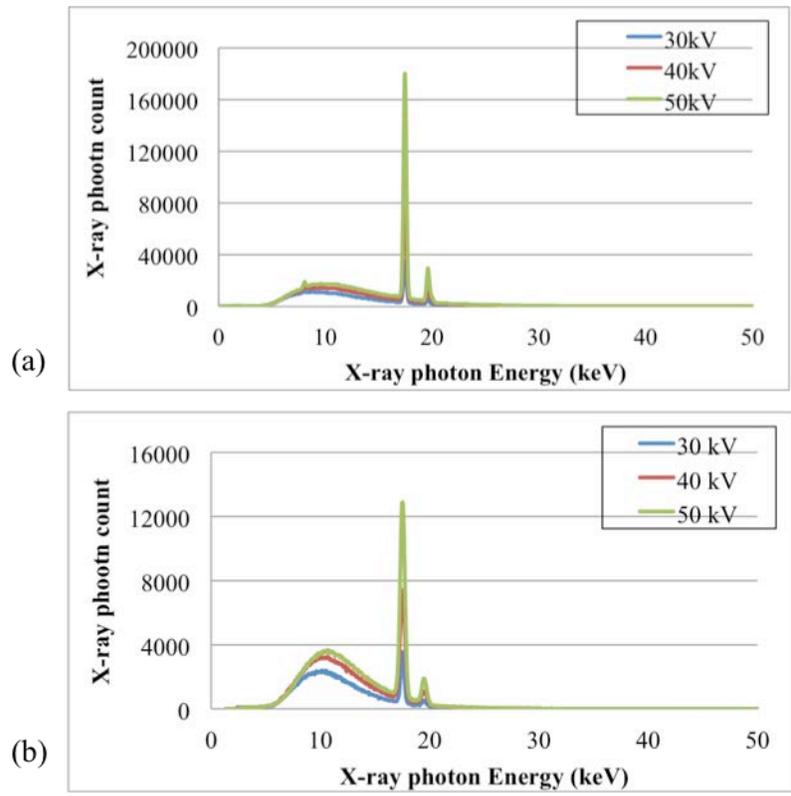

Figure 7

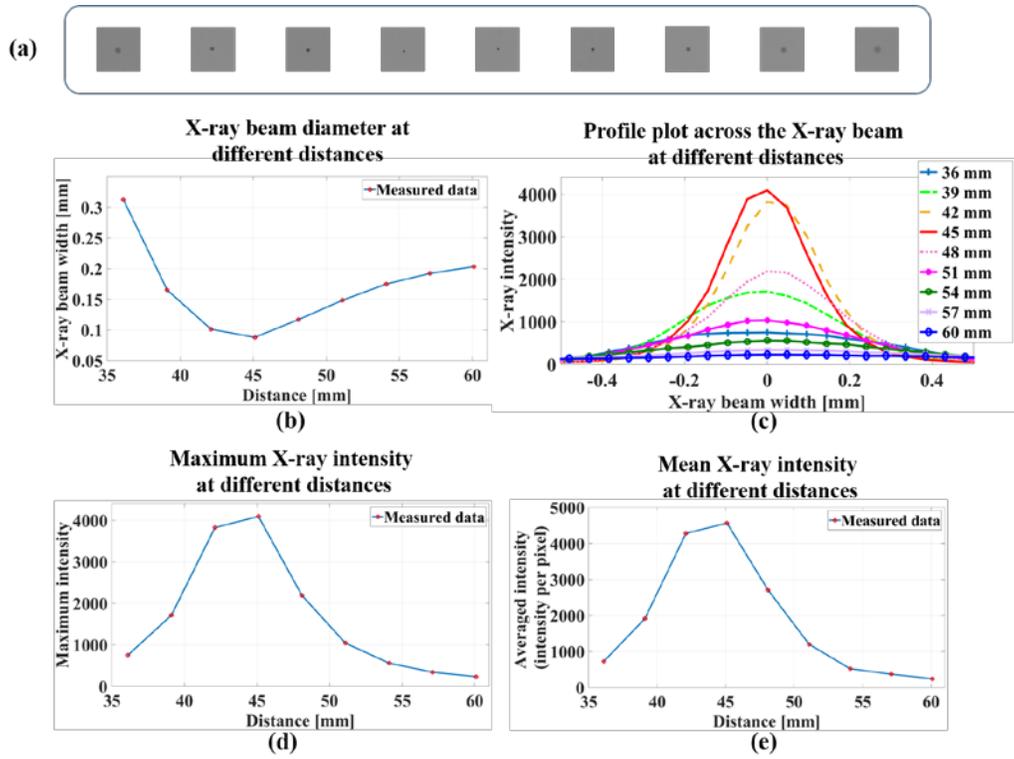

Figure 8

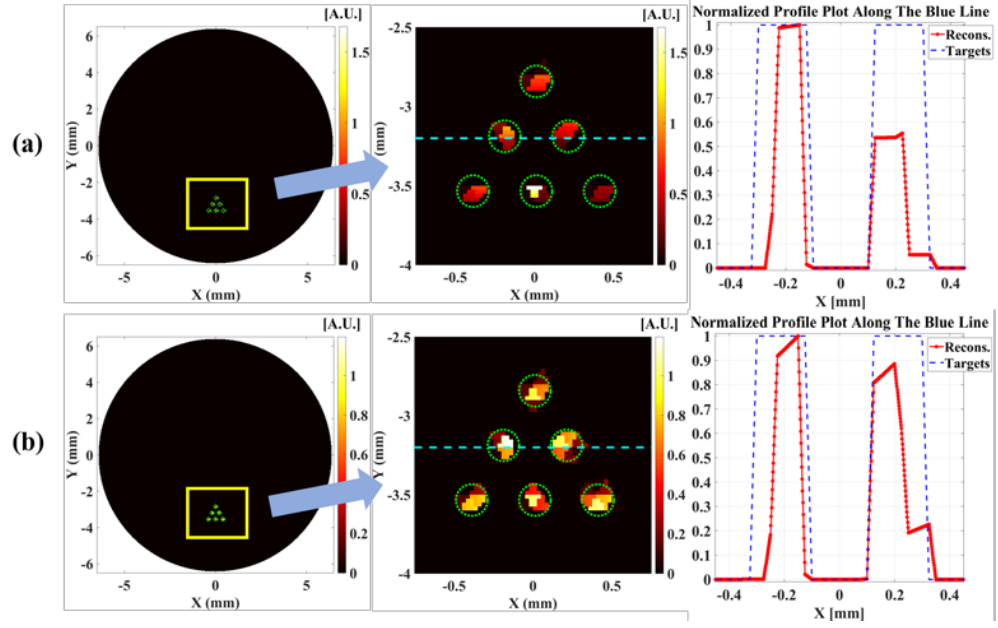

Figure 9

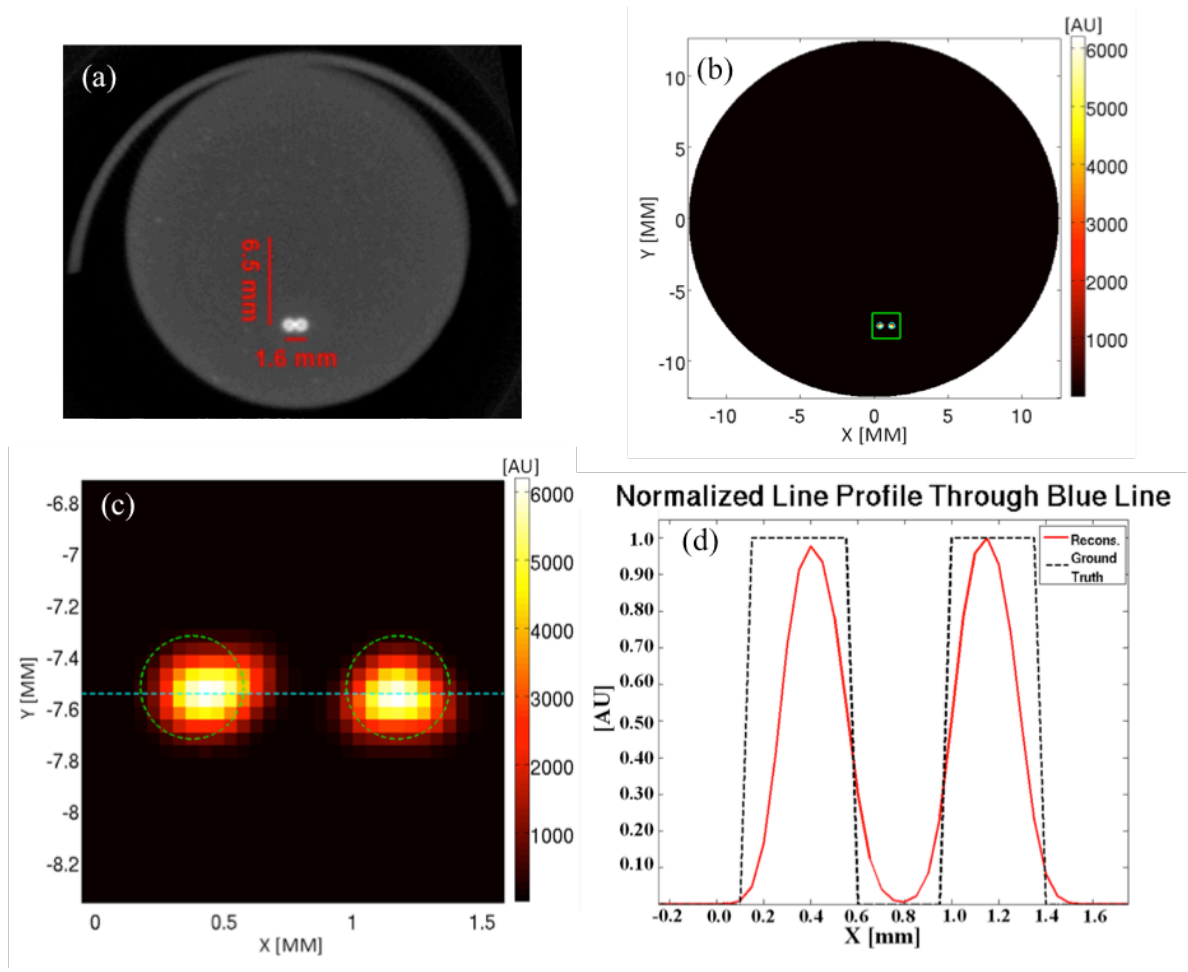

Figure 10

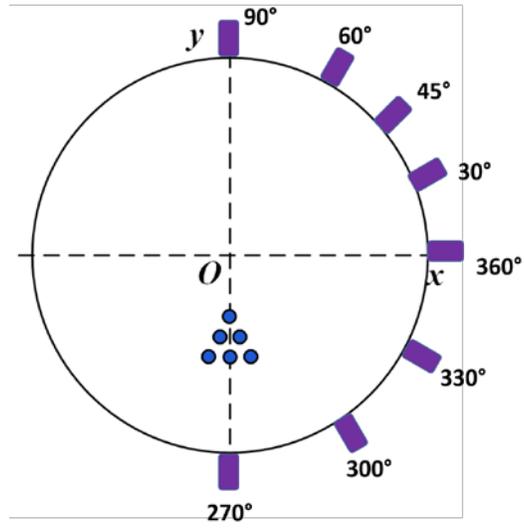

Figure 11

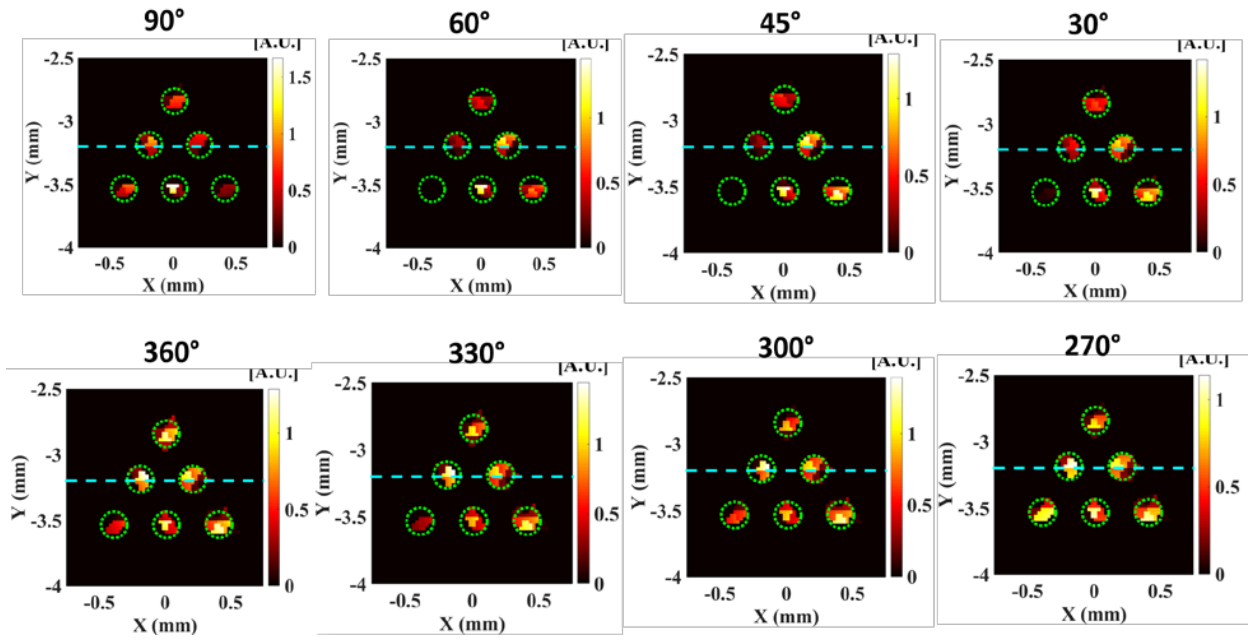

Figure 12

**Table 1**

| Number of Fiber Bundle | Diameter(mm)/TSE | CtCD(mm)/CDE | DICE |
|---|---|---|---|
| 1 | 0.1388/30.62% | 0.3713/7.19% | 47.11% |
| 6 | 0.1844/7.8% | 0.4156/3.9% | 41.86% |

**Table 2**

| Diameter (mm)/TSE | CtCD (mm)/CDE | DICE |
|---|---|---|
| 0.45/12.5% | 0.75/6.25% | 80.0% |